# Acoustically-driven surface and hyperbolic plasmon-phonon polaritons in graphene/h-BN heterostructures on piezoelectric substrates

R Fandan[1,2], J Pedrós[1,2], J Schiefele[3], A Boscá[1,2], J Martínez[1,4], and F Calle[1,2]

[1] Instituto de Sistemas Optoelectrónicos y Microtecnología, Universidad Politécnica de Madrid, Av. Complutense 30, Madrid 28040, Spain

[2] Departamento de Ingeniería Electrónica, E.T.S.I de Telecomunicación, Universidad Politécnica de Madrid, Av. Complutense 30, Madrid 28040, Spain

[3] Instituto de Ciencia de Materiales de Madrid, Consejo Superior de Investigaciones Científicas, C/ Sor Juana Inés de la Cruz 3, Madrid 28049, Spain

[4] Departamento de Ciencia de Materiales, E.T.S.I de Caminos, Canales y Puertos, Universidad Politécnica de Madrid, C/ Profesor Aranguren s/n, Madrid 28040, Spain

E-mail: rajveer.fandan@upm.es, j.pedros@upm.es

Abstract

Surface plasmon polaritons in graphene couple strongly to surface phonons in polar substrates leading to hybridized surface plasmon-phonon polaritons (SPPPs). We demonstrate that a surface acoustic wave (SAW) can be used to launch propagating SPPPs in graphene/h-BN heterostructures on a piezoelectric substrate like AlN, where the SAW-induced surface modulation acts as a dynamic diffraction grating. The efficiency of the light coupling is greatly enhanced by the introduction of the h-BN film as compared to the bare graphene/AlN system. The h-BN interlayer not only significantly changes the dispersion of the SPPPs but also enhances their lifetime. The strengthening of the SPPPs is shown to be related to both the higher carrier mobility induced in graphene and the coupling with h-BN and AlN surface phonons. In addition to surface phonons, hyperbolic phonons appear in the case of multilayer h-BN films leading to hybridized hyperbolic plasmon-phonon polaritons (HPPPs) that are also mediated by the SAW. These results pave the way for engineering SAW-based graphene/h-BN plasmonic devices and metamaterials covering the mid-IR to THz range.

## 1. Introduction

Surface plasmon polaritons (SPPs) are electromagnetic waves confined to the interface between two materials and accompanied by collective oscillations of surface charges. One of the most intriguing properties of SPPs is that their momentum is larger than that corresponding to free photons of the same frequency [1]. As a consequence, SPPs are bound modes whose fields decay exponentially away from the interface, exhibiting deep subwavelength confinement which results in strong light-matter interaction [2, 3]. Thus, SPPs in metals have attracted a strong interest over the last few decades for manipulating light and light-matter interactions at scales well beyond the diffraction limit [4] and have laid the foundation of a whole new range of fields including SPP-based nanophotonic devices [5, 6], metamaterials [7], metasurfaces [8], and quantum plasmonics [9]. However, the lifetime of SPPs in metals is extremely short when the light is confined to deep subwavelength scales and their properties cannot be modulated *in-situ* [10, 11]. In the case of doped graphene, SPPs can be confined to extreme subwavelength scales while preserving a long lifetime [10]. The carriers in graphene interact strongly with the surface optical (SO) phonons of polar substrates via the long-range Fröhlich coupling leading to hybridized surface plasmon-phonon polaritons (SPPPs) [12,13]. Moreover, unlike metals, the SPP (or SPPP) wavelength in graphene can be tuned *in-situ* through the modulation of the carrier density by electrostatic gating, thus providing a versatile plasmonic platform covering the mid-IR to THz range [14-16]. Hence, graphene plasmonics are being investigated for a large number of applications including two-dimensional transformation optics [17], optical signal processing [17, 18], single photon non-linear optics [19], biosensing [20], and integrated optics [21].

In order to excite a SPP (or SPPP) in graphene, a large momentum mismatch has to be overcome. Momentum can be gained either using near-field techniques [22-24], frequency mixing [25], or diffraction at nanostructures using far-field radiation, where the nanostructures can be made by patterning the graphene film itself [12, 26, 27] or the substrate where the graphene layer is transferred to [28, 29]. Surface acoustic waves (SAWs) have been shown to provide a suitable mechanism to launch propagating SPPPs in unpatterned graphene using simple far-field excitation [30]. An interdigital transducer (IDT) on a piezoelectric film is used to launch the SAW across the graphene sheet creating a tunable optical grating without the need of any patterning in either the graphene layer or the substrate, thus eliminating edge scattering and fragility issues. In addition to these advantages, the use of IDTs enables the fabrication of graphene plasmonic devices by the microelectronics industry, as compared to other schemes using an external mechanical vibrator [31]. Moreover, the use of IDTs permits to efficiently control the SPPP both temporally and spatially. Thus, the propagating SPPP can be switched electrically via the high-frequency signal at the IDT and its wavefront can be shaped by tailoring the IDT design. For example, curved IDTs creating interfering SAWs could be used for focusing the SPPP.

In this paper, we provide a complete simulation study of graphene SPPPs excited by means of a SAW in graphene/h-BN/AlN heterostructures with varying h-BN film thickness. The atomically flat surface without dangling bonds of h-BN is known to provide the largest carrier mobility in graphene as compared to any other insulating substrate [32-34]. We show that the introduction of an h-BN film between the graphene and the AlN substrate not only enhances the SPPP lifetime, but also significantly changes the hybridized SPPP dispersion, as compared to the previously studied graphene/AlN system [30]. The lifetime enhancement provided by the h-BN interlayer is shown to be related to both the higher carrier mobility in graphene and the larger lifetime of the surface phonons as compared to the AlN substrate. In addition, hyperbolic phonons appear in the case of multilayer h-BN films that also couple with the graphene carriers leading to hybridized hyperbolic plasmon-phonon polaritons (HPPPs) [24, 35]. We demonstrate that a SAW can also be used to couple light into HPPPs in the graphene/h-BN/AlN systems.

## 2. Optical properties of graphene, h-BN, and AlN

The frequency dependent conductivity of graphene in the local limit ($k \to 0$) and for $\omega \gg \tau_e^{-1}$ can be calculated using the Kubo formula [36] as

$$\sigma(\omega) = \frac{e^2(\omega + i\tau_e^{-1})}{i\pi\hbar^2} \left[ \frac{1}{(\omega + i\tau_e^{-1})^2} \int_0^\infty E \left( \frac{\partial F(E)}{\partial E} - \frac{\partial F(-E)}{\partial E} \right) dE + \int_0^\infty \frac{F(E) - F(-E)}{(\omega + i\tau_e^{-1})^2 - 4(E/\hbar)^2} dE \right], \quad (1)$$

where $F(E) = [exp\{(E - \mu_c)/k_B T\} + 1]^{-1}$ is the Fermi-Dirac distribution, $\mu_c$ is the chemical potential and $\tau_e$ is the electron relaxation time. $\tau_e$ is related to the mobility, $\mu$, and the Fermi energy, $E_F$, by $\tau_e = \mu E_F v_F^{-2}$ where $v_F$ is the Fermi velocity of $10^6$ ms$^{-1}$. The first and second terms in equation 1 represent the contributions of the intra-band ($\hbar\omega < 2E_F$) and inter-band ($\hbar\omega \geq 2E_F$) transitions, respectively, as shown in figure 1(a).

h-BN is an anisotropic van der Waals crystal with two IR active phonon modes: the out-of-plane ($\perp$) A$_{2u}$ phonon modes, with frequencies $\omega_{TO\,\perp} = 0.096$ eV and $\omega_{LO\,\perp} = 0.102$ eV, and the in-plane ($\parallel$) E$_{1u}$ phonon modes, with frequencies $\omega_{TO\,\parallel} = 0.169$ eV and $\omega_{LO\,\parallel} = 0.199$ eV [35]. The frequency dependent relative permittivity of h-BN is given by

$$\epsilon_m(\omega) = \epsilon_{\infty,m}\left(\frac{\omega_{LO\,m}^2 - \omega_{TO\,m}^2}{\omega_{TO\,m}^2 - i\gamma_m\omega - \omega^2}\right), \qquad (2)$$

where $m = \parallel, \perp$; $\epsilon_{\infty,\parallel} = 2.95$ and $\epsilon_{\infty,\perp} = 4.87$ are the high-frequency dielectric constants; and $\gamma_\parallel = 0.49$ meV and $\gamma_\perp = 0.62$ meV are the damping frequencies. The permittivity of h-BN becomes negative in the frequency range between the TO and LO phonons ($\omega_{TO\,m} < \omega < \omega_{LO\,m}$), leading to two reststrahlen bands. The opposite sign in the in-plane and out-of-plane components within these bands makes h-BN a hyperbolic material. The lower band presents type-I hyperbolicity ($Re(\epsilon_\parallel) < 0, Re(\epsilon_\perp) > 0$), whereas the upper band has type-II hyperbolicity ($Re(\epsilon_\parallel) > 0, Re(\epsilon_\perp) < 0$), as shown in figure 1(b).

AlN is only slightly anisotropic [37] as compared to h-BN. Thus, for simplicity, we assume an isotropic behaviour with a unique reststrahlen band located between $\omega_{TO} = 0.083$ eV and $\omega_{LO} = 0.111$ eV, as shown in figure 1(b). The frequency dependent relative permittivity of AlN is given by

$$\epsilon(\omega) = \epsilon_\infty + (\epsilon_0 - \epsilon_\infty)\left(\frac{\omega_{TO}^2}{\omega_{TO}^2 - i\gamma\omega - \omega^2}\right), \qquad (3)$$

where $\epsilon_0 = 7.37$ and $\epsilon_\infty = 3.93$ are the static and high-frequency dielectric constants, respectively, and $\gamma = 0.64$ meV is the damping frequency.

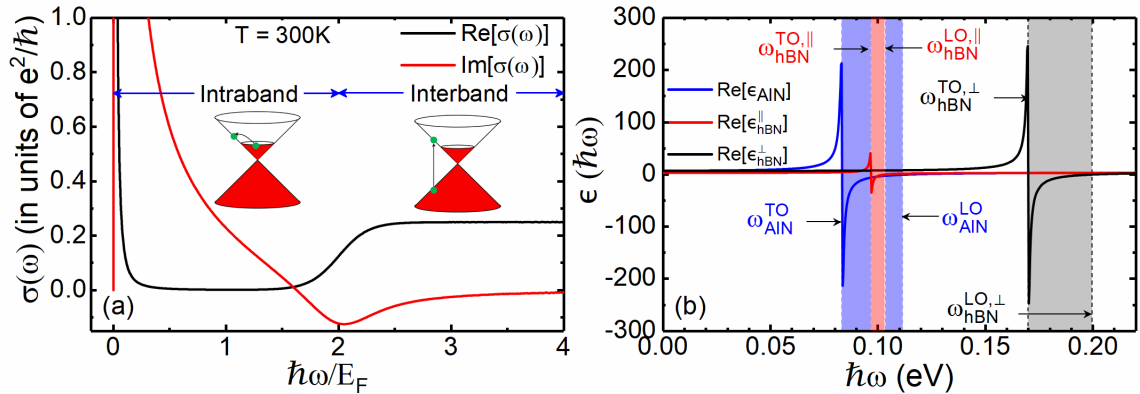

**Figure 1.** (a) Conductivity of graphene as a function of the photon energy normalized by the Fermi energy. The contribution from the intra- and inter-band transitions is indicated. (b) Permittivity of h-BN and AlN graphene as a function of the photon energy. Shaded regions are the reststrahlen bands, which appear in the frequency ranges between the TO and LO phonons.

## 3. Dispersion of SPPPs and HPPPs in graphene/h-BN/AlN systems

The carriers in graphene couple to the long-range electric field induced by optically active phonons in the surrounding materials. Both h-BN and AlN are polar materials, so that their SO phonons strongly couple to graphene carriers. Thus, graphene carriers can interact by exchange of SO phonons leading to a potential given by [34]

$$V_{\omega SO}(k,\omega) = \frac{4\pi\hbar\alpha c}{2k} \sum_{SO} \frac{e^{-2kz} a \omega_{SO}^2}{(\omega + i\tau_{SO}^{-1})^2 - \omega_{SO}^2}, \quad (4)$$

where $k$, $\omega_{SO}$, and $\tau_{SO}$ are the SO phonon wave vector, frequency, and lifetime, respectively, $z$ is the graphene-substrate separation, $a = (\epsilon_0 - \epsilon_\infty)/[(1+\epsilon_0)(1+\epsilon_\infty)]$, $\alpha = 1/137$ is the fine-structure constant, and $c$ is the velocity of the light in vacuum. $\tau_{SO}$ is estimated as the inverse of the damping rate of TO phonon, whereas $\omega_{SO}$ is the solution of the equation $\epsilon(\omega) + 1 = 0$.

Apart from the coupling to substrate phonons from the surrounding materials, carriers in graphene also couple to in-plane optical and out-of-plane (flexural) phonon modes in the graphene sheet itself. While the former would play a role only at plasmon energies above 0.2 eV [12], the latter is strongly suppressed in substrate supported graphene sheets [38, 39]. Hence, we have neglected here the coupling to both of these phonons.

The total effective carrier interaction results from Coulomb interaction, $V_c(k) = e^2/2k\epsilon_{vac}\epsilon_{eff}$ (with the permittivity of vacuum $\epsilon_{vac} = e^2/4\pi\hbar\alpha c$, and the effective relative permittivity given by the average of the background permittivity $\epsilon_{eff} = \left(1 + \epsilon_{AlN}^\infty + \left(\epsilon_{hBN}^{\infty,\perp} * \epsilon_{hBN}^{\infty,\parallel}\right)^{1/2}\right)/3$), and phonon exchange, $V_{\omega SO}(k,\omega)$. Within the random phase approximation (RPA), the total effective carrier interaction $V_{RPA}^{eff}(k,\omega)$ is given by [12, 13]

$$V_{RPA}^{eff}(k,\omega) = \frac{V_c(k)}{\epsilon_{RPA}} = \frac{V_c(k) + V_{\omega SO}(k,\omega)}{1 - (V_c(k) + V_{\omega SO}(k,\omega))\Pi_{\rho\rho}^0(k,\omega)}, \quad (5)$$

where $\epsilon_{RPA}$ is the total dielectric screening function and $\Pi_{\rho\rho}^0(k,\omega)$ is the non-interacting part (i.e. the pair-bubble diagram) of the charge-charge correlation function (2D polarizability). The latter is given by the modified Lindhard function [40, 41]

$$\Pi_{\rho\rho}^0(k,\omega) = -\frac{g_s g_v}{4\pi^2} \int d^2q \sum_{ss'} f_{ss'}(q,q+k) \frac{F(E_s(q)) - F(E_{s'}(|q+k|))}{E_s(q) - E_{s'}(|q+k|) + \hbar\omega + i\hbar\tau_e^{-1}}, \quad (6)$$

where $g_s = g_v = 2$ are the spin and valley degeneracies, respectively, $s, s' = \pm 1$ denote the band indices, $E_\pm(q) = \pm\hbar v_F q - \mu_c$ are the eigen energies, $F(E_s(q)) = [exp\{E_s(q)/k_B T\} + 1]^{-1}$ is the Fermi-Dirac distribution, and $f_{ss'}(q,q+k)$ is the band overlap of the wavefunction. The latter term includes the characteristic difference between the polarizability of the Dirac (massless) electron gas in graphene and a Fermi (massive) 2D electron gas in a semiconductor system. It is given by the expression

$$f_{ss'}(q,q+k) = \frac{1}{2}\left(1 + ss'\frac{q+k\cos\theta}{|q+k|}\right), \quad (7)$$

where θ is the angle between $q$ and $q+k$. In the limit $\omega > v_F k$ and $E_F \gg min(\hbar\omega, T)$, $\Pi_{\rho\rho}^0(k,\omega)$ can be approximated to

$$\Pi_{\rho\rho}^0(k,\omega) \approx \frac{E_F k^2}{\pi\hbar^2(\omega + i\tau_e^{-1})^2}. \quad (8)$$

The electron energy loss function $L(k,\omega) = -\text{Im}[1/\epsilon_{RPA}(k,\omega)]$ characterizes the spectral density of collective charge excitations. The SPPP dispersion can be obtained by solving the equation $\epsilon_{RPA}(k, \omega) = 0$, corresponding to the poles of $L(k, \omega)$. The lifetime of the SPPP $\tau_p$ can be calculated by solving the equation $\epsilon_{RPA}(k, \omega_p - i\tau_p^{-1}) = 0$.

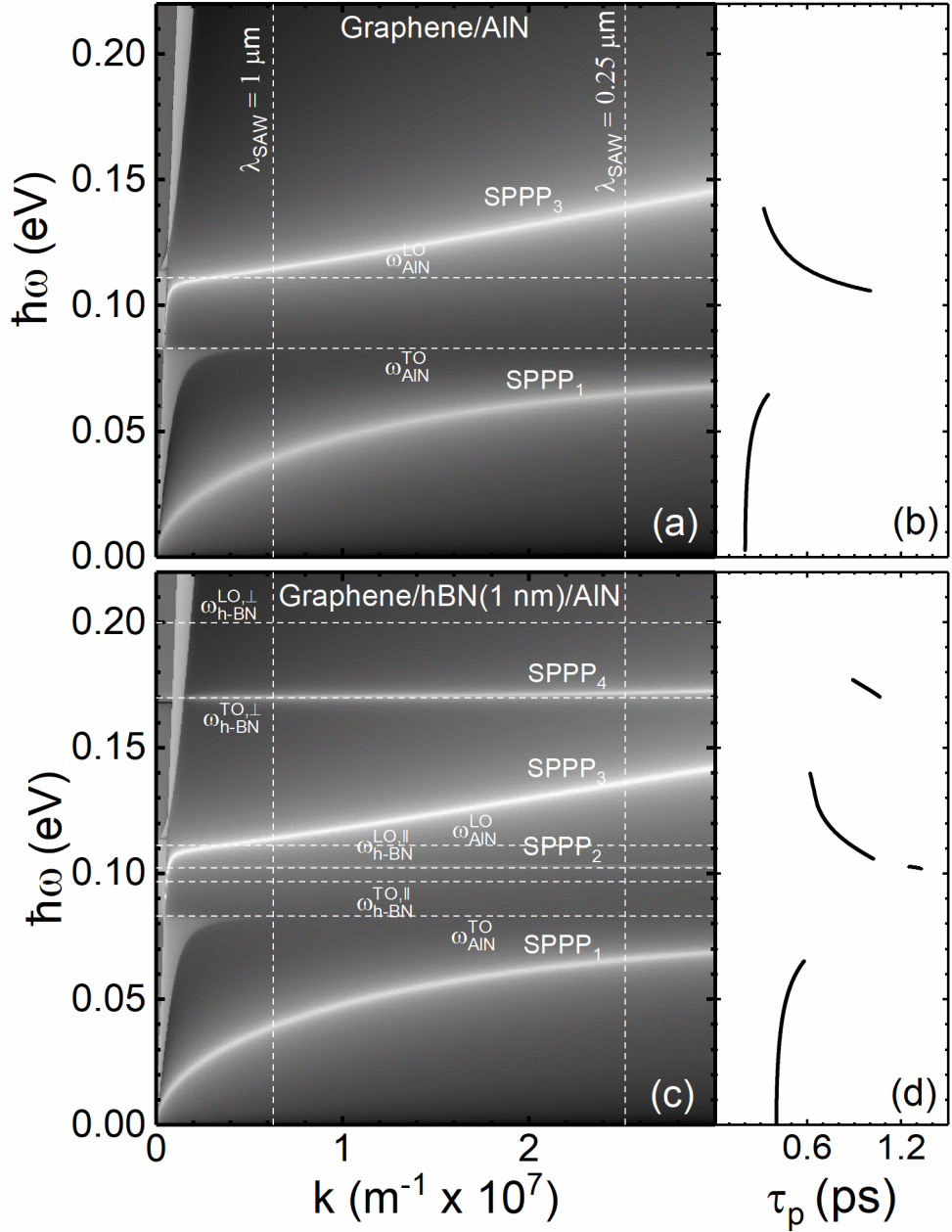

**Figure 2.** SPPP$_i$ dispersion for (a) graphene/AlN and (c) graphene/h-BN/AlN with an h-BN film thickness $d = 1$ nm. Parameters are $E_F = 0.4$ eV for both structures and $\mu = 5000$ and $10000$ cm$^2$ V$^{-1}$ s$^{-1}$ for graphene/AlN and graphene/hBN/AlN, respectively. Dashed horizontal lines correspond to the phonon frequencies of AlN and h-BN, whereas the dashed vertical lines indicate SAW wavelengths $\lambda_{SAW}$ of 0.25 and 1 μm. (b), (d) Lifetime of the SPPP$_i$.

Figures 2(a) and 2(c) depict the SPPP dispersion for the graphene/AlN and graphene/h-BN/AlN systems, where the latter has a 1 nm-thick h-BN interlayer. The graphene mobility has been considered to be enhanced by the h-BN film from $\mu = 5000$ to $10000$ cm$^2$ V$^{-1}$ s$^{-1}$, whereas $E_F = $

0.4 eV has been set for both cases. The dispersion of the graphene/h-BN/AlN system presents a larger number of SPPP$_i$ as compared to the graphene/AlN counterpart, since the electron-phonon interaction involves both h-BN and AlN surface phonons. In particular, SPPP$_1$ and SPPP$_3$ arise from the interaction of the graphene SPP with the SO phonon in AlN [34], as shown in figures 2(a) and 2(c), whereas the additional SPPP$_2$ and SPPP$_4$ in figure 2(c) arise, respectively, from the interaction with the ∥ and ⊥ SO phonons of h-BN. The SPPP$_2$ branch is weak in intensity and is very close to the LO phonon of lower reststrahlen band of h-BN. The hybridized SPPP$_i$ present different spectral weight and lifetime, which depend on the oscillator strength and phonon lifetime of the underlying materials. The lifetime $\tau_p$ of the SPPP$_i$ spans between the limits imposed by the graphene carrier relaxation time, $\tau_e$, and the SO phonon lifetime, $\tau_{SO}$. Since the h-BN film enhances the graphene carrier mobility, and hence $\tau_e$, the $\tau_p$ of SPPP$_1$ and SPPP$_3$ is increased. On the other hand, h-BN supports longer-lived SO phonons than AlN, providing SPPP$_2$ and SPPP$_4$ with the largest $\tau_p$ values (up to almost 1.4 ns for SPPP$_2$). Thus, the h-BN interlayer allows both to strengthen the SPPPs and to broaden the energy range that they cover.

The graphene/h-BN/AlN system permits to achieve deep subwavelength confinement of the radiation. Figure 3 presents the confinement ratio $\lambda_{air}/\lambda_{SPPP}$ as a function of the incident light wavelength $\lambda_{air}$. The mid-IR to THz range can be covered with confinement ratios of up to two orders of magnitude.

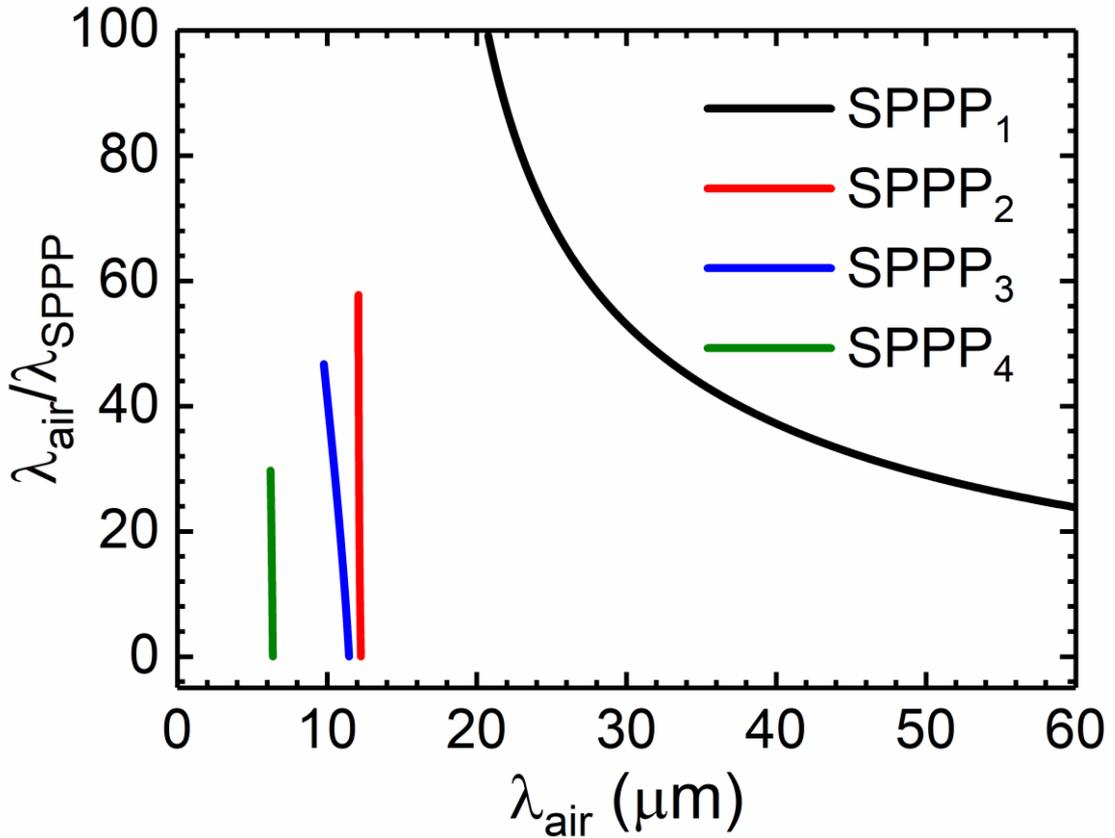

**Figure 3.** Confinement ratio $\lambda_{air}/\lambda_{SPPP}$ for the different SPPP$_i$ supported by the graphene/h-BN($d$ = 1 nm)/AlN system. Parameters are $E_F$ = 0.4 eV and $\mu$ = 10000 cm$^2$ V$^{-1}$ s$^{-1}$.

The dispersion of the graphene/h-BN/AlN system strongly depends on the thickness $d$ of the h-BN interlayer, as shown in figure 4. As $d$ increases slightly, SPPP$_1$ and SPPP$_4$ shift towards higher energies, whereas SPPP$_3$ shifts towards lower energies and SPPP$_2$ does not vary substantially, as shown in figures 2(c) and 4(a) and 4(b) for $d$ = 1, 5, and 10 nm, respectively. For $d$ = 50 nm, see figure 4(c), HPPPs [24, 40] appear within the h-BN reststrahlen bands as a result of the natural hyperbolicity of h-BN. These HPPPs appear also in the dispersion of the thinner multilayer h-BN films considered here but at higher $k$ values than those depicted in figures 2(b) and 4(a) and 4(b). In general, the thicker the h-BN interlayer, the stronger its influence on the dispersion curve of the graphene/h-BN/AlN system, with the HPPP$_i$ appearing at lower $k$ values, and the weaker the influence of the AlN susbstrate.

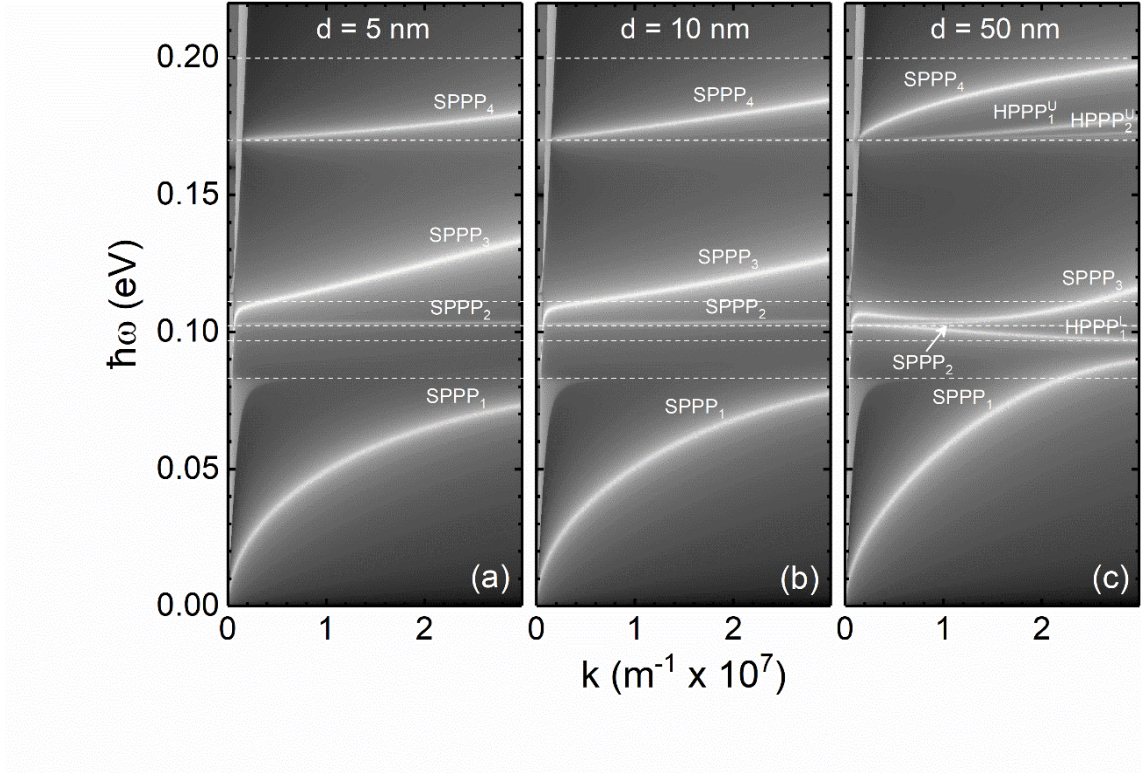

**Figure 4.** SPPP$_i$ and HPPP$_i^{U,L}$ dispersion for the graphene/h-BN/AlN system with h-BN film thickness $d$ of (a) 5 nm, (b) 10 nm, and (c) 50 nm. The superscripts in HPPP$_i^{U,L}$ denote the upper and lower reststrahlen bands. Parameters are $E_F$ = 0.4 eV and $\mu$ =10000 cm$^2$ V$^{-1}$ s$^{-1}$. Dashed horizontal lines correspond to the phonon frequencies of AlN and h-BN, as labelled in fig. 2.

SPPPs and HPPPs lead to enhanced absorption of the incident light. The transmittance and reflectance of the graphene/h-BN/AlN system can be calculated from the Fresnel reflection and transmission coefficients by means of the transfer matrix method. Hence, the transfer matrix is given by $M = B_{01} P B_{12}$, where $B_{01}$ and $B_{12}$ are the transfer matrices for the air/h-BN and h-BN/AlN interfaces, respectively, and $P$ is the propagation matrix. These matrices are given by the expressions

$$B_{01} = \begin{pmatrix} 1 + \sigma(\omega)/Z_1 + Z_0/Z_1 & 1 - \sigma(\omega)/Z_1 - Z_0/Z_1 \\ 1 + \sigma(\omega)/Z_1 - Z_0/Z_1 & 1 - \sigma(\omega)/Z_1 + Z_0/Z_1 \end{pmatrix} \quad (9.1)$$

$$B_{12} = \begin{pmatrix} 1 + Z_1/Z_2 & 1 - Z_1/Z_2 \\ 1 - Z_1/Z_2 & 1 + Z_1/Z_2 \end{pmatrix} \quad (9.2)$$

$$P = \begin{pmatrix} e^{-ik_1^z d} & 0 \\ 0 & e^{ik_1^z d} \end{pmatrix}, \tag{9.3}$$

where $j = 0,1,2$ for air, h-BN, and AlN respectively, $Z_j = \omega \varepsilon_j^\perp (k_j^z)^{-1}$ is the impedance of $j$th material, where $k_j^z = \left[\varepsilon_j^\perp\left((\omega/c)^2 - k^2/\varepsilon_j^\parallel\right)\right]^{-1/2}$, and $\sigma(\omega)$ is the conductivity of graphene defined in equation 1.

For TM polarized light, the Fresnel reflection ($r_{TM}$) and transmission ($t_{TM}$) coefficients are given in terms of elements of the transfer matrix $M$ as $r_{TM} = M_{21}/M_{11}$ and $t_{TM} = 1/M_{11}$. The absorption of the light is given by Im[$r_{TM}$], which is interpreted as the contour plot of electron energy loss function in figure 2.

## 4. SAW-mediated SPPPs and HPPPs

The wave vector mismatch between the light and the SPPPs and HPPPs in the graphene/h-BN/AlN system can be overcome using an IDT on the surface of the AlN substrate to launch a SAW propagating across the graphene/h-BN heterostructure. The SAW produces a sinusoidal deformation of the surface creating a virtual dynamic diffraction grating that provides the extra wave vector required for the light to couple into the SPPPs [34] and HPPPs. Thus, in the presence of a SAW of wavelength $\lambda_{SAW}$ and amplitude $\delta$, the transmittance of TM-polarized light is reduced as the SAW-induced diffraction grating scatters the incident light with wave vector $(k_{\parallel 0}, k_{z0})$ into the various diffraction orders $(k_{\parallel m}, k_{zm})$ with

$$k_{\parallel m} = (\omega/c)\sin\theta + m2\pi/\lambda_{SAW}, \tag{10}$$

$$k_{zm} = \sqrt{(\omega/c)^2 - k_{\parallel m}^2}, \tag{11}$$

where $\theta$ is the angle of off-normal incidence and $m$ is an integer. The intensity of the diffracted light in $m$th order, normalized to the intensity of the incoming beam, is given by

$$I^m = \left| r_{TM} J_m(2k_{z0}\delta) \left[1 + \frac{m\pi}{\lambda_{SAW} k_{zm}} \frac{1+r_{TM}}{r_{TM}}\right] \right|^2, \tag{12}$$

where $J_m$ is the Bessel function for the $m$th diffraction order.

Figure 5 plots the extinction spectrum $1-T^{TM}(\delta)/T^{TM}(0)$ of the graphene/AlN and graphene/h-BN($d$ = 1nm)/AlN systems for the first diffraction order ($m = 1$) induced by a SAW with $\delta = 4$ nm and $\lambda_{SAW} = 250$ nm. As described in the previous section, the SPPP dispersion of the graphene/AlN system is strongly modified by the introduction of a thin h-BN interlayer. This is also reflected into the extinction spectra, as shown in figures 5(a) and 5(b) for the SPPP$_1$ and SPPP$_3$, respectively. In addition to a frequency shift, the enhancement of the mobility induced by the h-BN allows SPPPs to live longer and propagate longer distances. Furthermore, new SPPPs appear in the extinction spectra, as shown for example in figures 5(c) for the SPPP$_4$.

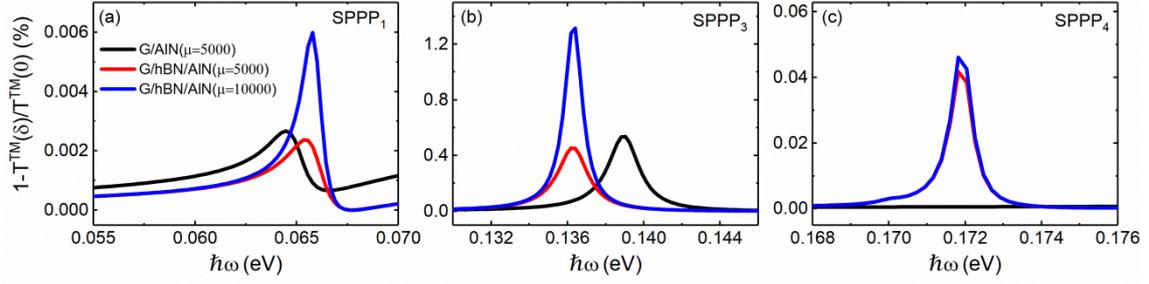

**Figure 5.** Extinction spectra of (a) SPPP$_1$, (b) SPPP$_3$, and (c) SPPP$_4$ for the graphene/AlN and graphene/h-BN($d$ = 1 nm)/AlN systems in the presence of a SAW ($m$ = 1, $\delta$ = 4 nm, and $\lambda_{SAW}$ = 250 nm) for variable carrier mobility $\mu$ (in units of cm$^2$ V$^{-1}$ s$^{-1}$) and $E_F$ = 0.4 eV.

In addition to the tunability of the SPPP and HPPP dispersion in the graphene/h-BN/AlN system provided by the thickness of the h-BN interlayer, graphene presents an intrinsic degree of tunability by means of the modulation of its carrier density $n$, or $E_F = \hbar v_F \sqrt{\pi n}$, through electrostatic gating. The effect of these two tuning mechanisms on the extinction spectra of the graphene/h-BN/AlN system for the first diffraction order ($m$ = 1) induced by a SAW with $\delta$ = 4 nm and $\lambda_{SAW}$ = 250 nm is shown in figure 6. Rows in figure 6 show the effect of varying the h-BN film thickness for different SPPP$_i$ at fixed $E_F$ values, whereas columns show the effect of varying $E_F$ for the same SPPP$_i$ and h-BN film thickness. In general, as $E_F$ increases, all SPPP$_i$ peaks in the extinction spectra get more intense in magnitude. However, the peak intensity of particular SPPP$_i$ is especially enhanced at certain h-BN film thickness and $E_F$ combinations. Figure 6(d) illustrates an example of this enhancement for the SPPP$_1$ in the graphene/h-BN($d$ = 50 nm)/AlN system at $E_F$ = 0.4 eV. Similarly, the SPPP$_3$ gets strongly enhanced in the graphene/h-BN(d = 10 nm)/AlN system at $E_F$ = 0.2 eV, figure 6(b), and in the graphene/h-BN(d = 50 nm)/AlN system at $E_F$ = 0.4 eV, figure 6(e). Therefore, a detailed calculation of the dispersion of the graphene/h-BN/AlN system, both in terms of h-BN thickness as well as graphene doping, is required for the design of the most efficient SAW-mediated plasmonic devices.

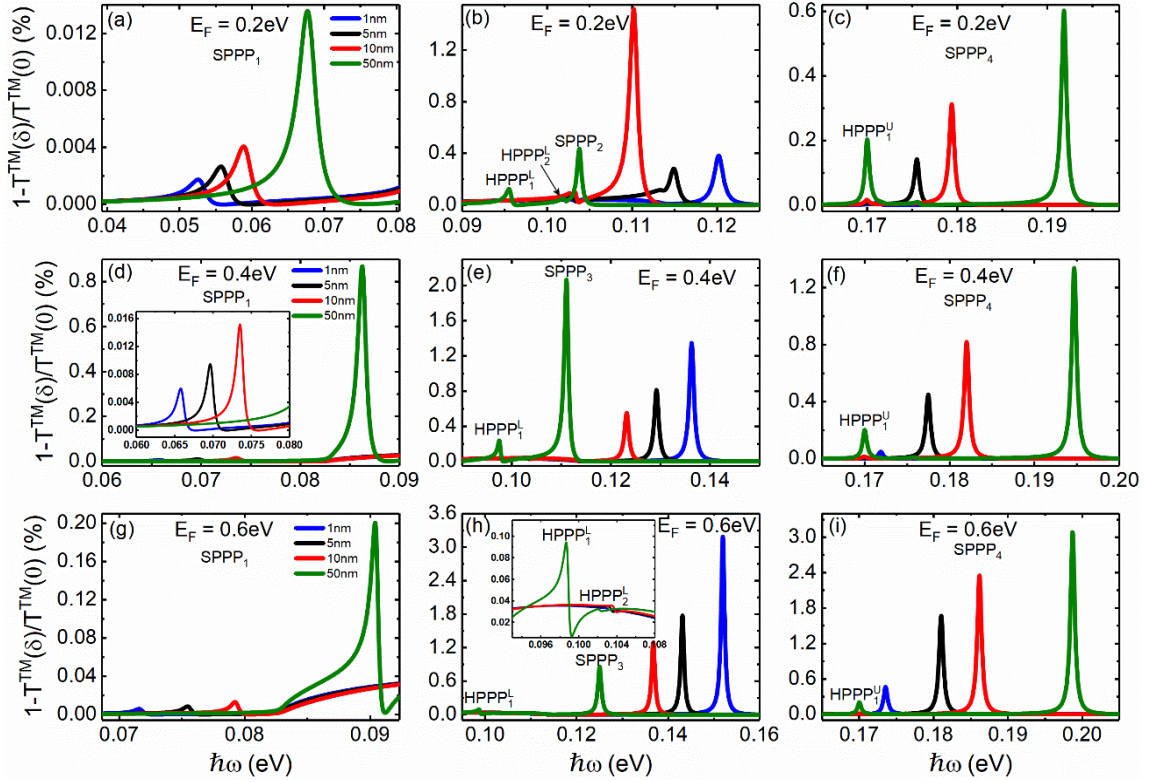

**Figure 6.** Extinction spectra of (a), (d), (f) SPPP$_1$, (b), (e), (h) SPPP$_2$ and SPPP$_3$ and (c), (f), (i) SPPP$_4$ for graphene/h-BN/AlN systems with carrier mobility $\mu = 10000$ cm$^2$ V$^{-1}$ s$^{-1}$ and various h-BN film thickness $d$ in the presence of a SAW ($m = 1$, $\delta = 4$ nm, and $\lambda_{SAW} = 250$ nm). The Fermi energy is $E_F = 0.2$ (upper panel), 0.4 (intermediate panel), and 0.6 eV (lower panel). Insets in (d) and (h) show magnified views. Additional HPPP$_i$ labelled for discrimination.

## 5. Conclusions

In conclusion, we have theoretically demonstrated that SAWs can be used to generate SPPPs and HPPPs in graphene/h-BN heterostructures on a piezoelectric substrate, such as AlN, where the SAW creates a dynamic virtual diffraction overcoming the wave vector mismatch of SPPPs and HPPPs with the incident light. Graphene electrons couple to surface phonons of both the h-BN interlayer and the AlN substrate providing a complex SPPP dispersion relation that strongly depends on the thickness of the h-BN film. In addition, hyperbolic phonons appear in the case of multilayer h-BN films that also couple with the graphene carriers leading to HPPPs. The h-BN interlayer is demonstrated to not only significantly change the hybridized SPPP dispersion but also to enhance the SPPP lifetime, as compared to the simpler graphene/AlN system. The lifetime enhancement provided by the h-BN film is shown to be related to both the higher carrier mobility induced in graphene and the larger lifetime of the h-BN surface phonons. Therefore, the graphene/h-BN/AlN system provides an advantageous platform with greater plasmon robustness and tunability for future SAW-based plasmonic devices.

**Acknowledgments**

The authors thank Fernando Sols and Francisco Guinea for helpful discussions. This work has received funding from the European Union's Horizon 2020 Research and Innovation Programme under Marie Skłodowska-Curie Grant Agreement No 642688, the NANOGREAT Network of Infrastructure of the European Institute of Innovation and Technology - Raw Materials, and from the Spanish MINECO through project GRAFAGEN (ENE2013-47904-C3). J.P. acknowledges financial support from MINECO (Grant RyC-2015-18968).

**ORCID iDs**

**R Fandan** ORCID 0000-0002-4885-8853

**F Calle** ORCID 0000-0001-7869-6704